\documentclass[runningheads]{llncs}

\usepackage[T1]{fontenc}
\usepackage{graphicx}
\usepackage{algorithm}
\usepackage{algpseudocode}
\usepackage{multirow}
\usepackage{enumitem}
\usepackage{optidef}
\usepackage{amsmath,amssymb}
\usepackage{mathtools}
\usepackage{pgfplots}
\usepackage{pgfplotstable}
\usepackage{subcaption}

\DeclarePairedDelimiter\ceil{\lceil}{\rceil}

\newcommand\norm[1]{\left\lVert#1\right\rVert}
\newcommand{\probP}{\text{I\kern-0.15em P}}

\begin{document}
\title{UpLIF: An Updatable Self-Tuning Learned Index Framework}
%
%
\author{Alireza Heidari\orcidID{0009-0002-0414-7360} \and
Amirhossein Ahmadi\orcidID{0009-0002-9563-0912} \and
Wei Zhang\orcidID{0009-0006-2866-8688}}
%

%
\institute{
Huawei Cloud\\
\email{\{alireza.heidarikhazaei,amirhossein.ahmadi,wei.zhang6\}@huawei.com}}
\maketitle              
\begin{abstract}
The emergence of learned indexes has caused a paradigm shift in our perception of indexing by considering indexes as predictive models that estimate keys' positions within a data set, resulting in notable improvements in key search efficiency and index size reduction; however, a significant challenge inherent in learned index modeling is its constrained support for update operations, necessitated by the requirement for a fixed distribution of records. Previous studies have proposed various approaches to address this issue with the drawback of high overhead due to multiple model retraining. In this paper, we present UpLIF, an adaptive self-tuning learned index that adjusts the model to accommodate incoming updates, predicts the distribution of updates for performance improvement, and optimizes its index structure using reinforcement learning. We also introduce the concept of balanced model adjustment, which determines the model's inherent properties (i.e. bias and variance), enabling the integration of these factors into the existing index model without the need for retraining with new data. Our comprehensive experiments show that the system surpasses state-of-the-art indexing solutions (both traditional and ML-based), achieving an increase in throughput of up to $3.12 \times$ with $1000 \times$ less memory usage.

\end{abstract}

\section{Introduction}

\noindent
\textbf{Context.} Machine Learning has been progressively adopted to improve index systems, referred to as 	\textit{learned indexes}, due to its swift predictive abilities and capacity to encapsulate the distribution patterns of records~\cite{heidari2019holodetect,heidari2024record}. The recursive model index (RMI) \cite{kraska2018case} proposed an immutable ML model that, given a key, identifies the range where the corresponding data is located. This approach rivals or even suppresses classical data indexing systems (such as B+Tree) in search performance~\cite{10.14778/3551793.3551823}. In this modeling, the position of each record is estimated by calculating the value of the given key in a monotonic function which can be approximated based on a Cumulative Distributed Function (CDF)~\cite{heidari2020sampling,livshits2020approximate,10.14778/3587136.3587148}. However, the drawback of learned index modeling is that the distribution of records must remain fixed, which poses challenges for operations that update the key domain (such as \texttt{Insert} and \texttt{Delete}). Since the integration over the CDF domain must always sum to $one$, any records' insertion or deletion, necessitates a non-local update to the current learned model.

To overcome the challenge of an updatable learned index, three main approaches have been proposed~\cite{10.14778/3594512.3594528}: \textit{(i) Delta-buffer, (ii) In-place, (iii) Hybrid structures}. The \textit{delta-buffer} solutions involve storing incoming records in a buffer when multiple data records are assigned to the same position to postpone the required updates to the current learned model. However, when the buffer size exceeds a certain threshold, buffer pruning occurs and the data merge into the model, as suggested by LIPP~\cite{wu2021updatable}. In contrast, the \textit{in-place} approaches reserve empty placeholders in the original key domain for unseen records. The problem with these solutions, such as Alex~\cite{ding2020alex} is that once the predicted offset by the learned model is filled, subsequent updates are placed in the next available slots regardless of key sequence, leading to extensive search in key lookup. Finally, the \textit{hybrid} approach combines placeholders and buffers for entries that do not fit in the key domain, aiming to strike a balance between lookup efficiency and update speed for different system workloads. DILI~\cite{10.14778/3598581.3598593}, a state-of-the-art hybrid solution, employs a tree structure that facilitates level-by-level lookups using the models and buffered data stored in the nodes. The buffered data influence the tree height, leading to an increase in the tree height as more updates are processed. 

\vspace{3pt}
\noindent
\textbf{Challenges.} The existing solutions for updatable learned indexes exhibit the following limitations~\cite{10.14778/3551793.3551848}: \textit{(1) \textbf{Ignoring the Training Cost}.} The primary issue with earlier methods is the need to modify the learned model by keeping updates or constructing several models. Retraining in the simplest scenario (that is, using a linear regression model) requires $O(N)$, and a cascading effect through their update structures costs $O(N log N)$. \textit{(2) \textbf{Constrained to Specific Incoming Update Distribution}.} These approaches strongly consider the specific spread of the incoming data and existing keys in the key domain. When they encounter a discrepancy in the incoming distribution, their throughput drops drastically. \textit{(3) \textbf{Lack of Generality}.} There is no method to mitigate the effects of various algorithms on the learned index. These methods consistently use the same training algorithm, such as linear regression, throughout the system's lifetime.

\vspace{3pt}
\noindent
\textbf{Approach.} In this paper, we present \textit{\textbf{UpLIF}}, an adaptive self-tuning learned index framework, that \textit{adjusts the learned index model on the incoming updates to postpone the model retraining and utilizes the distribution of the updates for its model adjustment.} UpLIF uses the hybrid approach and introduces \textbf{\underline{B}}alanced \textbf{\underline{M}}odel \textbf{\underline{A}}djustment \textbf{\underline{T}}ree (\textit{\textbf{BMAT}}) for its delta-buffer. BMAT captures incoming updates to the base model and calculates a linear adjustment for the learned index model approximation error at query time. UpLIF also learns the distribution of the incoming updates in the runtime to add in-place placeholders in the key domain. 

UpLIF also introduces a reinforcement learning-based optimization to adjust the BMAT structure to ensure high performance and minimal memory usage. This agent assesses performance metrics, including the height of BMAT, and takes the following actions to improve the structure of BMAT: \textit{ (1)} training on a data subset to lower the height of BMAT, or \textit{ (2)} transitioning between different types of BMAT that are more suitable according to the existing workload and data volume.

UpLIF shows notable enhancements over a wide range of indexing techniques: our approach can increase performance by up to $3.12 \times$ compared to other indexing methods (both traditional and machine learning-based) while requiring up to $1000 \times$ less memory for its index structure. UpLIF also addresses the following technical challenges: \textit{(1)~\textbf{Balancing}.} Learning a linear model with $C$ features and $N$ training data requires $O(CN)$. UpLIF delays retraining to keep updated operating cost low by employing BMAT. However, balancing in BMAT can be disrupted by incoming update traffic. To address this, UpLIF uses logarithmic operations to rebalance and eliminate the effects of update distribution change.
\textit{(2)~\textbf{Movement}.} Inserting a key in the middle of a list involves dividing the list into two sublists at the insertion point. This process requires relocating one of the sub-lists to a different memory location, leading to a complexity of $O(N)$. UpLIF reduces the relocation numbers by introducing a gap size. It also expands the relocation range by placing placeholders by forecasting the upcoming updates on the relocation gap based on the incoming update distribution.
\textit{(3)~\textbf{Self-Tuning}.} Update operations lead to an increase in the BMAT height, thus increasing the search cost. UpLIF autonomously identifies the optimal balance point to maintain system performance. It decides to reduce tree height based on different \textit{performance signals} by solving an optimization problem in which the cost function aims to maximize system throughput while simultaneously minimizing memory usage. 
\textit{(4)~\textbf{Generic}.} UpLIF has generic modeling for modification of the index tree that is compatible with any kind of key space (e.g., numerical, string) and can work with any learned index model type.

\vspace{3pt}
\noindent
\textbf{Contributions.} The main contributions of this work are as follows:
\begin{enumerate}[leftmargin=*]
\item \textit{Developing a modular architecture independent of the underlying learned index model (Section~\ref{frameworkOverview}).} This framework accepts any input type for which a comparison measure can be defined within that input space.


\item \textit{Presenting BMAT, a balanced tree structure for storing data needed for model adjustment (Section~\ref{sec:model-adjust}).} This structure controls the index adjustment inherent properties (i.e., bias and variance). It captures updates to the base model at query time and uses a linear correction for the learned index model approximation error. 

\item \textit{Developing an optimization agent adaptively tunes the framework to maintain the performance high and index size low (Section~\ref{sec:signals}}). This agent employs reinforcement learning to identify the BMAT states require tuning and selects the most suitable tuning action to optimize the overall index system.

\item\textit{Evaluating UpLIF against various state-of-the-art indexing systems (Section~\ref{sec:eval}).} UpLIF exhibits higher performance among all tested systems through our extensive experiments while maintaining significantly lower index sizes.

\end{enumerate}

\section{Framework Overview}
\label{frameworkOverview}
In this section, we explain the constitutional modules of the UpLIF architecture (Figure~\ref{fig:framework}). UpLIF begins by initializing the model $M_0\in \mathbb{M}$ from the sorted array $keys$, where $\mathbb{M}$ is a hypothesis space. Then it takes lookup or update queries with respect to a key $k$. Once the system is initialized, it accepts queries that look up or modify a specific key, $k$. To answer these queries, UpLIF uses the following four core modules:

\begin{figure}
     \centering
     \begin{subfigure}[t]{0.44\textwidth}
         \centering
         \includegraphics[width=\textwidth]{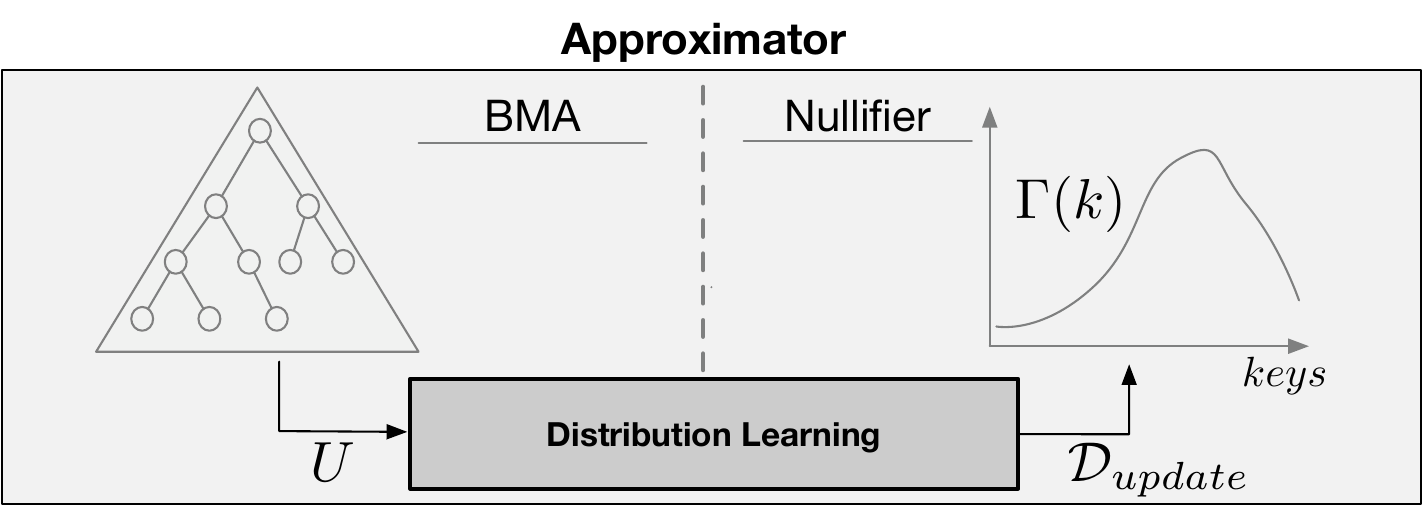}
         \caption{}
         \label{fig:approx}
         \label{}
     \end{subfigure}
     \begin{subfigure}[t]{0.55\textwidth}
         \centering
         \includegraphics[width=\textwidth]{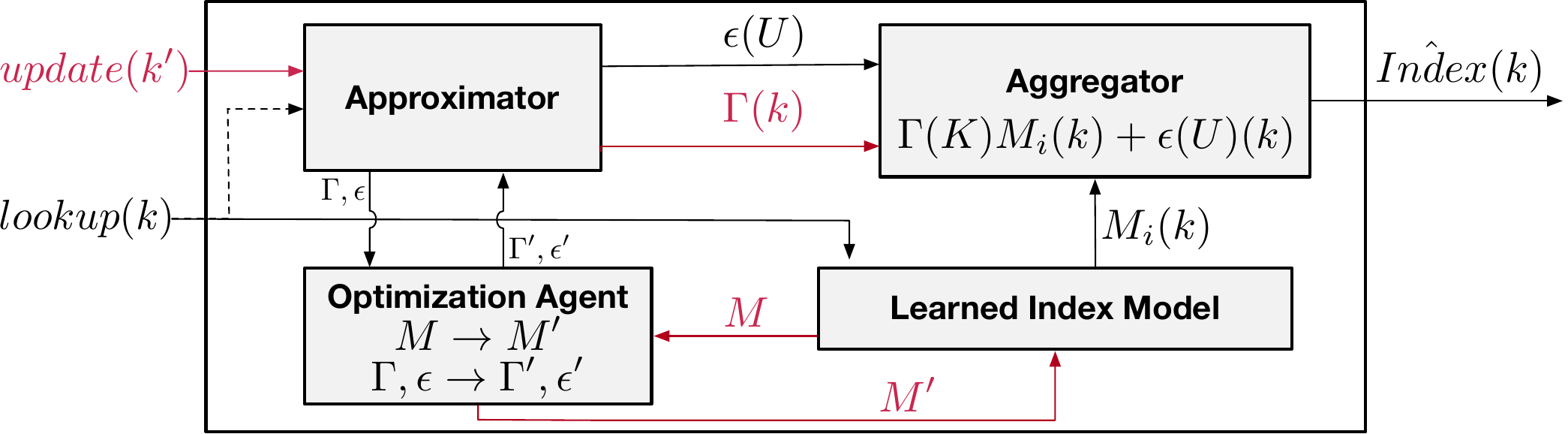}
         \caption{}
         \label{fig:framework}
     \end{subfigure}
      \caption{\small{} (a) The approximator module demonstrates that trains the distribution $\mathcal{D}_{update}$ from the incoming updates $U$. (b) Overview of the UpLIF's modules. Module 1 predicts an approximate index based on the existing model $M_i(k)$. Module 2 then uses probabilistic methods to generate an offset to adjust the approximate index adjustments, denoted $\epsilon_U$, as well as an adjustment for the model error, denoted $\epsilon_M$. Module 3 combines the results of Module 1 and Module 2 with the original model $M_i(k)$ to create a new model for lookup queries for a given key $k$. Finally, Module 4 evaluates the system's performance and, utilizing updates, the current model $M$, variance $\Gamma$, and bias $r$, along with a Q-learning trained model, produces a new model $M'$ or updates to $\Gamma'$ and $r'$. }

\end{figure}




\noindent\textbf{Module 1: Learned Index Model.} This module is initialized with the model $M_0$ learned on the initial sorted keys. The only requirement for choosing a learned index algorithm is that it must maintain the data in a sorted order. This is due to: \textit{(i)} making the data in order causes the inefficiency in the range queries, and \textit{(ii)} the operation of the Approximator module relies on the data being sorted. Consequently, methods such as Alex~\cite{ding2020alex},
which alter the key sequences, cannot be used as the learned index base model in our framework. In our evaluations (Section~\ref{sec:eval}), we use RadixSpline~\cite{kipf2020radixspline} as the underlying learned index model.

\vspace{3pt}\noindent\textbf{Module 2: Approximator.} This module estimates the index adjustment caused by incremental updates that have not yet been applied to the model in Module 1. 
We introduce the concept of Balanced Model Adjustment for the Approximator to accept the incoming updates and control the index model bias and variance for index model correction in a logarithmic time. The model correction that occurs in the approximation module makes the retraining unnecessary to answer queries (Sections \ref{sec:index-model-appr} and \ref{sec:bmat}).

\vspace{3pt}\noindent\textbf{Module 3: Aggregator.} This module combines the results of the \textit{Learned Index Model} and \textit{Approximator} modules and produces a range for the last mile search. Moreover, it performs a logarithmic search in the result range and returns the index of the query as the result. Aggregator also handles the placement of appropriate placeholders in the key space based on the update distribution to ensure efficient future insertions by optimizing space allocation for new entries (Section~\ref{sec:nullifier}).

\vspace{3pt}\noindent\textbf{Module 4: Optimization Agent.} Accepting millions of updates without system tuning influences UpLIF's performance and memory usage. The Optimization Module considers various performance metrics that impact the system's throughput and index size. Then, it uses reinforcement learning to adaptively tune the index system structure. This module switches between BMAT types (i.e. Red-Black Tree and B+Tree). It also reduces the height of the BMAT by retraining on the subset of the data. To avoid training the model on a large amount of data, the module always divides the key domain and selects the smallest training set heuristically (Section~\ref{sec:signals}).

\section{Model Adjustment Overview}
\label{sec:model-adjust}
In this section, we explain our technique to adjust the index model based on the incoming updates to postpone retraining. First, we introduce an index model approximation that is able to solve the model adjustment without requiring model retraining. Then, we present the Balanced Model Adjustment Tree (BMAT) and Nullifier that help to solve the approximation problem.

\subsection{Preliminaries}

\newcommand{\tabincell}[2]{\begin{tabular}{@{}#1@{}}#2\end{tabular}}

\begin{table}[t]
    \centering
    \caption{Notations}\label{tab:notations}
    \begin{tabular}{c|p{0.6\columnwidth}}
        \hline 
        \tabincell{l}{$D$} & \tabincell{l}{Initial data in the system.} \\ \hline
         \tabincell{l}{$\mathcal{D}_{update}$} &\tabincell{l}{The distribution of incoming updates}  \\ \hline
        \tabincell{l}{$U$} & \tabincell{l}{Set of new updates drawn from $\mathcal{D}_{update}$} \\ \hline
        \tabincell{l}{$M(.)$} & \tabincell{l}{Initial learned index model trained on $D$} \\ \hline
        \tabincell{l}{$E$} & \tabincell{l}{The estimation error of $M(.)$} \\ \hline
         \tabincell{l}{$M'(.)$} &\tabincell{l}{The model retrained model of $M$.}  \\ \hline
          \tabincell{l}{$K$} &\tabincell{l}{The size of the moving records in memory}  \\ \hline
         \tabincell{l}{$\Gamma(.)$} &\tabincell{l}{Scalier function (Def. \ref{def:scaler}) over $keys$ }  \\ \hline
         \tabincell{l}{$\mathbb{M}$} &\tabincell{l}{Learned index model hypothesis space}  \\ \hline
         \tabincell{l}{$r(.)$} &\tabincell{l}{Bias function (Def. \ref{bareDef}) over $keys$}  \\ \hline
         \tabincell{l}{$d_{MAX}$} &\tabincell{l}{Maximum gap between continuous keys}  \\ \hline
         \tabincell{l}{$\alpha$} &\tabincell{l}{Average of of $\Gamma(.)$}  \\ \hline
         \tabincell{l}{$\xi$} &\tabincell{l}{Index approximation margin}  \\ \hline
         \tabincell{l}{$L$} &\tabincell{l}{BMAT (Section~\ref{sec:bmat}) height}  \\ \hline
         \tabincell{l}{$\delta$} &\tabincell{l}{Estimation confidence level}  \\ \hline
    \end{tabular}
\end{table}

\begin{center}
\begin{figure}[t]
  \centering
  \makebox[\columnwidth][c]{\includegraphics[width=0.7\columnwidth]{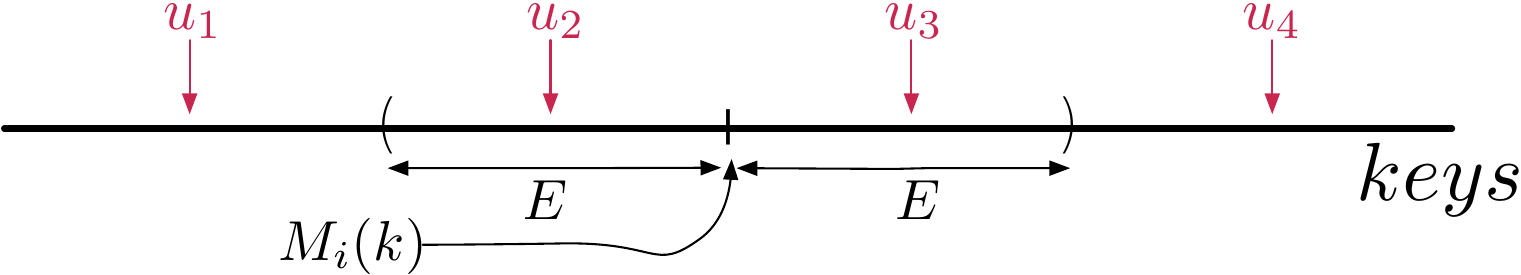}}
\vspace{-2em}
  \caption{The learned index model $M_i$, characterized by a nonzero error $E$, maps four ranges over the key space. Incoming updates, treated as a random variable, affect this range in four distinct ways: $u_1$ uniformly shifts all elements without changing the range size; $u_2$ expands the range on the left by shifting $M_i(k)$ rightward; $u_3$ enlarges the range on the right; and $u_4$ neither alters the error range size nor moves any internal components.}
  \label{fig:update_pattern}
\end{figure}
\end{center}

In this section, we provide the necessary definitions to formally state the problem. Table~\ref{tab:notations} denotes the notation used in this investigation. A learned index model $M(.)$ is a type of index system that uses machine learning algorithms to predict the range of the index $M(key)\pm E$ based on the provided $key$. To propose our framework for updating the learned index model, we must introduce the fundamental idea of updating the domain of a monotonically increasing function. Update queries (\texttt{INSERT} and \texttt{DELETE} operations) can alter the domain of $M(.)$, leading to a residual error (as shown in Figure~\ref{fig:update_pattern}). By keeping track of these errors, we can apply them to the model output, as if the model were re-trained on the entire updated data. The definition~\ref{def:goal} puts this in a formal linguistic way.

\begin{definition}
    Let $D_0$ be the initial data and $U_0$ be the first batch of update queries. Let $M_0=M(D_0)$ be the model trained in $D_0$ and $M_1=M(D_0\cup U_0)$ be the model trained in the updated data $D_1=D_0\cup U_0$. Let $\epsilon_0(.)$ be an error approximator for the first batch of updates. Our goal is $M(D_0\cup U_0) \approx M(D_0) + \epsilon_0(U_0) ~~~~or~~~~ M_1 = M_0 + \epsilon_0(U_0)$.  In general, 
    \begin{equation} 
        M_{i+1} = M_i + \epsilon_i(U_i),
       \label{bareDef}
    \end{equation}
    where $i$ denotes the batch number of updates and $\epsilon_i(U_i)$ is the error of the corresponding batch.
    \label{def:goal}
\end{definition}
This modeling suggests that if we can identify the set of incoming updates for the set $U_i$, we can use this to anticipate modifications to the model that has been trained on the updated data. To properly define this model, certain preliminary steps must be taken. Modeling $\epsilon_i(U_i)$ suggests a parallel way to describe index calculation, as the two parts of the updated model do not depend on the result of the other part.  

\begin{definition}
    Suppose that $f(x)$ be a \textit{continuous} and \textit{monotonic} increasing function, $g(x)$ called \textit{$t(x)$-expanded range} of $f(x)$ iff 
        \begin{equation} 
        g'(x) = t(x)f'(x)
    \end{equation} 
    which shows the relation between derivatives of $g(x)$ and $f(x)$. In the most basic scenario, when $t(x)$ is constant, the range of $g$ increases linearly. For instance, $f(x) = 4x$ is a 2-expanded version of $g(x) = 2x$, meaning that for every increment in $x$, the function $f$ occupies double the range compared to $g$.
    \label{expandDef}
\end{definition}

\begin{definition} [Scalier]
    $scalier_{s}$ is an operator that takes a function $f$ as input and scales it point-wise at the domain intersection of $f$ and $s$, in other words,
    \begin{equation} 
        scalier_{s}(f) =      \begin{cases}
       s(x)f(x) &\quad x\in Domain(f)\cap Domain(s)\\
       f(x) &\quad \text{otherwise.} \\ 
     \end{cases}
    \end{equation}
    \label{def:scaler}
\end{definition}

\begin{definition}[Nullify]
    Nullify is an operator that operates on a list of keys $L$. Introduces gaps between the elements of the list based on a given maximum distance $d_{MAX}$ and a distribution $\mathcal{D}$. In essence, this operator can be seen as a scalar transformation $scaler_\mathcal{D}$ on the index function that produces the list $L$.
    \label{def:nuliffy}
\end{definition}

\subsection{Index Model Approximation}
\label{sec:index-model-appr}
Upon receiving an update, the index function is modified so that all keys that exceed the updated value increase by one unit, while the index function remains unchanged for values less than the lead update received. Following updates $U$, each element $k$ in the original dataset $D$ is adjusted by the count of updates $x$ in $U$ where $x < k$. Initially, the index model is denoted as $M$. Once the comprehensive set of possible updates $\mathcal{U}$, drawn from the distribution $\mathcal{D}_{update}$, is applied, this index model evolves into a modified version $M_{\mathcal{U}}$, symbolized as $M^\star$. The goal of the optimization problem is to identify a model $M'$ that reduces the following cost function by considering a subset of updates $U\subseteq \mathcal{U}$:

\begin{equation}
\begin{aligned}
\mathcal{L}_{M'} = \sum\limits_{k\in D\cup U} \norm{M^\star(k)-M'(k)}
\end{aligned}
\label{eq:cost}
\end{equation}

\noindent therefore, our optimization problem can be defined as;
\begin{equation}
\resizebox{.9\hsize}{!}{$
\begin{aligned}
\min_{M'\in \mathbb{M}} \quad & \mathcal{L}_{M'}\\
\textrm{s.t.} \quad & \mathop{\mathbb{E}}_{u\sim \mathcal{D}_{update}}[M']=\mathop{\mathbb{E}}_{u\sim \mathcal{D}_{update}}[M^\star]\\
&  \probP\big[ \vert M'(k)-M'(u) \vert<\frac{1}{2}\big] \leq \tau \quad \forall u\sim \mathcal{D}_{update}, k\in D\cup U\\
\end{aligned}
\label{eq:opt}
$}
\end{equation}

\noindent Where $\mathbb{M}$ denotes the hypothesis space of the models and $\tau$ is a parameter chosen by the user to specify the size of the gap in the optimization solution. It is crucial to understand that determining the smallest $\tau$ renders the problem NP-Hard, requiring an exhaustive search across the whole $\mathbb{M}$ to identify the model that optimally predicts the gap for the unseen data $u \in \mathcal{D}_{update}$. It is essential to recognize that errors are inherent in every index model. Adding a bias value maintains the distance between keys constant, yet scaling a function alters the spacing between values within the function's range, thereby adjusting the estimator's variance. In system index estimation, it is crucial to prioritize an unbiased estimator despite the risk of higher variance, as increased variance results in a wider search range during the final stages, whereas bias compromises the reliability of the outcome. Searches typically progress from the center towards the boundaries, following normal distribution curves. Given that the likelihood of locating the target value in the distribution's tail is low, we strive to keep the optimization problem (Equation~\ref{eq:opt}) unbiased. This may lead to increased system variance; however, with lengthy tails, such variance is less harmful compared to the bias. It also considers the distribution of incoming updates; it is crucial to consider the interval linked to the most probable incoming updates, as indicated by the second optimization condition.

\vspace{3pt}
\noindent\textbf{Optimization Relaxation.} Addressing the optimization problem \ref{eq:opt} in its general form is exceedingly challenging, thus we introduce a relaxation for the function shape $M'$, representing it as a transformation from the original index function $M$. Various methods exist to transform a function, with the most straightforward which is the linear transformation, which involves using a coefficient and an additive function. A function ascends or descends along the vertical axis through the addition or subtraction of a constant, referred to as a bias. Furthermore, multiplying by numbers greater than one causes the vertical axis values to expand, whereas multiplication by numbers less than one results in a contraction of these values. These dynamics are encapsulated in linear approximation by the line's slope and the bias.

In this study, to simplify the optimization challenge described for Equation~\ref{eq:opt},  we introduce the class $\mathfrak{C}$ for $\mathbb{M}$ as linear combinations of the index function, expressed by $M'(k) = \Gamma(k)M(k) + r(k)$. Here, $\Gamma(k)$ acts as the Scalier (Definition~\ref{def:scaler}) for $M(k)$, and $r(k)$ accounts for the influence of bias error (Definition~\ref{bareDef}). 
Also from Definition \ref{expandDef}, $M'$ is approximately $\frac{d\Gamma(k)}{dk}$-expanded range of $M$. This linear structure allows us to assign specific roles to each segment of $\Gamma(k)$ and $r(k)$, thereby breaking down the optimization into two separate phases:

\vspace{3pt}
\noindent\textbf{Phase 1.} During this phase, the optimizer adjusts $r(k)$ to ensure the model is unbiased concerning the existing data in $D \cup U$, using the current model $M(k)$ and scalier $\Gamma(k)$.

\noindent\textbf{Phase 2.} In this stage, the optimizer modifies $\Gamma(k)$ to ensure the model remains unbiased towards the unseen data within distribution $\mathcal{D}_{update}$ based on the existing model $M(k)$ and bias function $r(k)$. Initially, given the unknown nature of $\mathcal{D}_{update}$, a uniform distribution is assumed. This is later adapted to an observation likelihood as system knowledge increases, thereby generating additional vacant areas in crucial domains where updates are more likely to occur, creating optimal spaces for future system updates.

\noindent While \textit{Phase 2} operates independently on the incoming updates and estimates $\mathcal{D}_{update}$, \textit{Phase 1} involves a step function with $|U|$ steps, each occurring at a unique position within the domain of keys. Constructing this function is feasible, yet updating the index function with each new update is necessary as the system may receive lookup queries. Therefore, a rapid method to execute these phases is essential. In the following section, we introduce a submodular framework that can approximately address the optimization challenge by altering local properties.

\vspace{-6pt}
\subsection{Balanced Model Adjustment Tree}
\label{sec:bmat}
In this section, we introduce \textbf{\underline{B}}alanced \textbf{\underline{M}}odel \textbf{\underline{A}}djustment \textbf{\underline{T}}ree (\textbf{BMAT}), a tree-shaped structure provides a deterministic approximation based on the distribution of the incoming update $\mathcal{D}_{update}$. By traversing BMAT, UpLIF determines the parameters needed for the adjustment of the model (that is, $\Gamma(k)$, $r(k)$, and \textit{hyperparameters}). Therefore, it guarantees that we achieve the performance of the learned index model with an additional logarithmic cost to identify the required adjustment. We implement the concept of BMAT using two types of balanced trees: \textit{(1) Red-Black Tree (RBMAT)}, and \textit{(2) B+Tree (B+MAT)}. Each of these types of BMAT is suitable for different sizes of datasets and operations, as we discuss in Section~\ref{sec:tune} (see Figure~\ref{fig:treecmp}).

BMAT acts as a delta-buffer structure construct in a subset $U^G\subseteq U$ that cannot be accommodated in the index structure. It offers a time complexity of $O(\log |U^G|)$ and a space complexity of $O(|U^G|)$. Each node in BMAT holds a fixed amount of data, and the structure is balanced to maintain optimal tree height and prevent performance decline. There are two possibilities for a key lookup: \textit{(1)}~If the key $k$ is present in $U$ that could not fit in a position of the index system, then the value of the query is already stored in a node, and \textit{(2)}~If $k$ is not found in $U$, then the number of children on the left side of each node on the subtree of the target node can determine the update bias value $r(k)$, which is indeed an effective method for 	\textit{Phase 1}. This deterministic structure introduces a forecast for the upcoming updates to address \textit{Phase 2} and allocates new empty slots for future updates, thus expanding the scope of the last-mile search.

\begin{figure}[]
  \centering
  \makebox[\columnwidth][c]{\includegraphics[width=0.9\columnwidth]{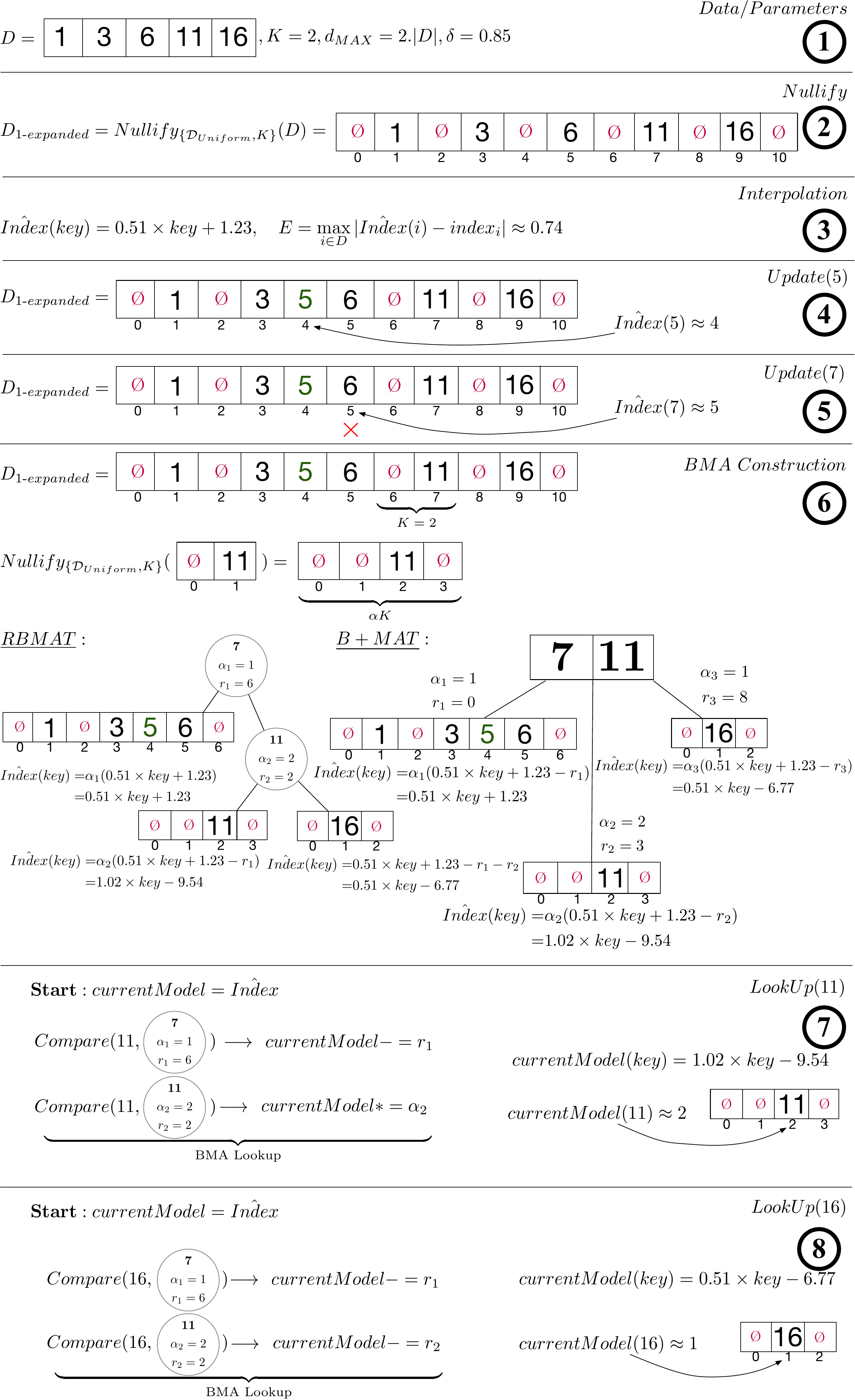}}
  

\caption{\small UpLIF overview on update and lookup operations. \textit{(1)} Initially, UpLIF creates a placeholder structure in the key domain and constructs an index model on top of it. \textit{(2)} An update with $key=5$ arrives, and it can be placed in the empty placeholder. (3) An update with $key=7$ arrives that makes conflict with $key=6$ using the current model. UpLIF divides the key domain into three segments and adjusts the previous model for each segment without retraining. It also adds placeholders with $\alpha=2$ in the middle segment (Section~\ref{sec:nullifier}). Finally, two nodes are added to BMAT that can be a red-black tree (left) or a B+Tree (right).}
  \label{fig:example}
\end{figure}

For the key $k$ update, a lookup occurs first, if the key is in the BMAT, the value is updated. If the key in the index structure has an empty place, its value is inserted in the empty place. Otherwise, the node is broken into three segments. UpLIF employs a hyperparameter $K$, which denotes the number of elements following the key index $k$, known as $aux$. The pair $k$ and $aux$ constitute the middle segment. This segment is then expanded through Nullifier (Section~\ref{sec:nullifier}). Then, the key $k$ and $aux$ (a key $K$ element after the key $k$) are added to the BMAT as nodes and point to the left, middle, and right segments with their updated bias values. Figure~\ref{fig:example} shows an example of the update operations in BMAT.

\subsection{Update Placeholder (\textit{Nullifier})}
\label{sec:nullifier}
When managing data accompanied by a workload $\mathcal{D}_{keys}$, it is essential to have an in-place system that distributes space between them to accommodate incoming updates based on the approximation of the updates. This is the role of \textit{Nullifier} (Figure~\ref{fig:approx}), which operates on a distribution $\mathcal{D}_{keys}$ in $domain$, a value $d_{MAX}$ as the maximum gap, and input data $D=[k_1,k_2,\dots,k_N]$ drawn from $\mathcal{D}_{keys}$, and then produces a $\mathcal{D}_{update}$-expanded (see Definition~\ref{expandDef}) of $D\sim \mathcal{D}_{keys}$. This involves calculating the number of gaps between keys $k_i$ and $k_j$ for $i<j$,
\begin{equation}
\label{eq:nullifier}
    GapSize(k_i,k_j) = \ceil*{\frac{d_{MAX}.\int_{k_i}^{k_j}\mathcal{D}_{update}(x)dx}{\int_{k_1}^{k_N}\mathcal{D}_{update}(x)dx}} 
\end{equation}

\noindent
To approximate the distribution $\mathcal{D}_{update}$ with incoming data $U$, we use a Gaussian Mixture Model (GMM).
Nullifier creates gaps in $D$ by selecting all successive elements and using Equation \ref{eq:nullifier}. For simplicity, when $j=i+1$, it is denoted by $\Gamma(k)=GapSize(k_{prev},k)$. Subsequently, according to Definition \ref{def:nuliffy}, produces a series of \texttt{NULL} values for each $k\in key$. The updated index for each $k$ is calculated as $GapSize(k_1,k)\times Index(k)$, with $\Gamma(k)$ vacant slots preceding it and $\Gamma(k_{next})$ vacant slots following it.

Note that the denominator is constant and only needs to be computed once. Then, we create a new vector using the values in $D$ and the calculated gap between its components, which we designated as \texttt{NULL}. To calculate the expansion of the key $k$, we determine the distance between the first key and $k$, as well as the number of elements less than $k$. This is equal to $\sum_{k': k'<k} \Gamma(k') + \vert \{k'\in D: k'<k \}\vert$. Instead of utilizing this formula inefficiently, which involves multiplying the index function for scaling, we opt to average these gaps and utilize a constant value instead.
\begin{equation}
\label{eq:alpha_avg}
    \alpha = \frac{\sum_{k\in D} \Gamma(k)}{N}
\end{equation}

\section{Adaptive System Tuning}
\label{sec:signals}
In this section, we present the optimization technique we use in UpLIF to self-tune its structure to maintain high performance and low memory usage. First, we cover the performance measures that influence the performance of UpLIF and index memory size. Then, we show the possible system tuning that can be applied to the index structure. Then, we present our designed reinforcement learning-based optimization agent that considers the performance measures and tuning actions and adjusts UpLIF's index structure with the goal of: \textit{(1)} Increasing the system throughput and \textit{(2)} Decreasing the index memory footprint. 

\subsection{Performance Measures}
\label{sec:pm}
Below are the performance measures that affect overall system performance and memory usage. Our optimizer must consider these metrics for system tuning.

\noindent
\textbf{Balanced Structure Height.}
The most crucial performance measure is the BMAT height. When the height of the index tree increases relative to the size of the processor caches, cache misses increase and cause a performance drop for BMAT traversal. By incorporating this measure into UpLIF, we can mitigate the decline in performance and achieve better overall system performance.

\noindent
\textbf{Granularity Measure.} This performance measure is the size of the effective segment range when creating a new node for BMAT. If the size gets smaller, the workload is skewed into a small range of keys increasing the system latency for update operations.

\noindent
\textbf{Error Scaling.} This factor evaluates the influence of $\Gamma(k)$ on the variation error of the model. As discussed in Section \ref{sec:index-model-appr}, the scaling factor $\alpha>1$ can increase the last-mile search, which affects the system query performance for lookup operations.

\noindent
\textbf{Number of the Models.} This performance measure keeps track of the number of distinct models stored on BMAT. The greater number of active models in the system shows higher memory usage for the indexing.

\begin{figure}[t]
  \centering
  \makebox[\columnwidth][c]{\includegraphics[width=0.6\columnwidth]{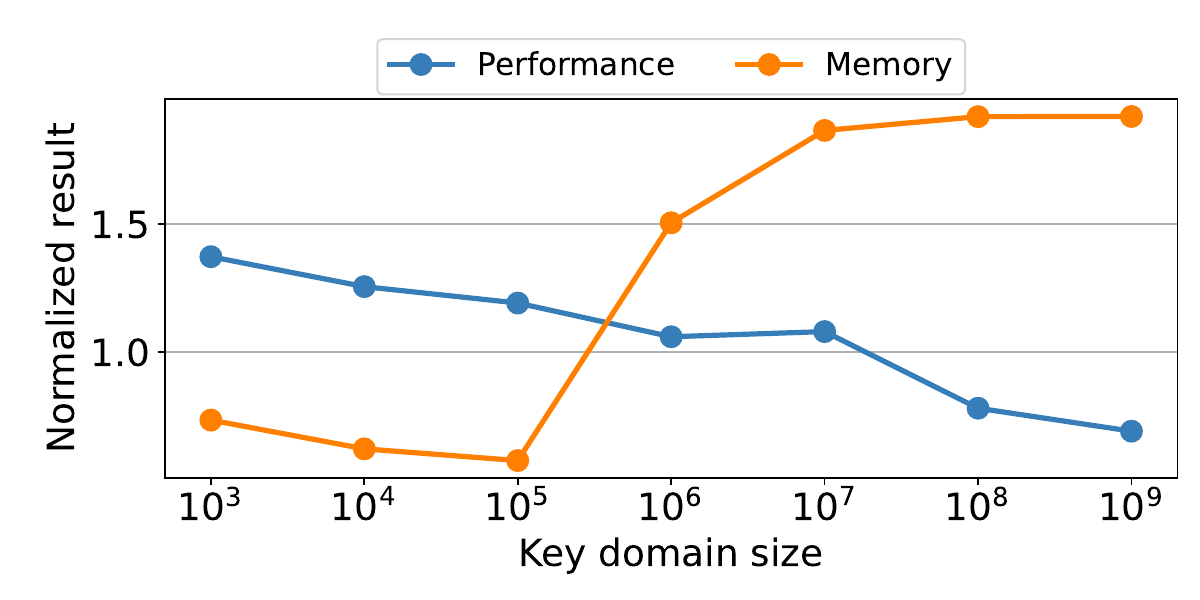}}
  \vspace{-1em}
  \setlength{\belowcaptionskip}{-2em}
  \caption{\small{} Comparison between various BMAT types. The plots show the performance and memory consumption of RBMAT normalized to B+MAT. Lower is better for Memory and higher is better for performance.}
  \label{fig:treecmp}
\end{figure}

\subsection{System Tuning Actions}
\label{sec:tune}
UpLIF system tuning can be performed in two different ways: \textit{(1)} Transitioning between various BMAT types and \textit{(2)} Reducing the BMAT height by retraining on the subset of keys. These adjustments in the BMAT structure influence the overall performance and memory consumption of UpLIF.

\noindent
\textbf{Transition between BMAT Types.}
We develop BMAT based on two different balanced trees: Red-Black Tree (RBMAT) and B+Tree (B+MAT). 
RBMAT is a binary tree and differs from B+MAT in that keys and pointers are clustered in memory; therefore, we obtain efficient cache behavior on both disk and in memory~\cite{heidari2024metahive}. However, for small data sizes, RBMAT is more efficient, as it does not have lookup overhead at each node, and while the BMAT traversal path can be kept in the cache, RBMAT performs better. 
In terms of memory, BMAT has the overhead of empty places in each node and RBMAT has the overhead of pointers to the right, left, and parent nodes for each node. RBMAT has a smaller size compared to B+MAT if the B+MAT nodes are empty. However, if the B+MAT nodes are mostly filled, RBMAT gets higher memory consumption to keep incoming updates. Figure~\ref{fig:treecmp} shows the comparison between the performance and memory consumption of RBMAT and B+MAT on a read-heavy workload (see Section~\ref{sec:eval}). The results show RBMAT data normalized against B+MAT data. In our setup, we observed that until $100$K keys, RBMAT acts quite better than B+MAT as it has higher throughput and lower index size. However, after this point, B+MAT performs better and is the ideal solution for the BMAT type.

\vspace{3pt}
\noindent
\textbf{Retraining on the Subset of Data.}
Another type of system tuning on BMAT is retraining on the subset of the data to reduce BMAT height. The growth in tree height due to our model approximation mechanism necessitates a periodic reverse process to prune the BMAT structure, maintaining the performance of our framework. We introduced a greedy approach to select the minimum number of nodes in both RBMAT and B+MAT to have minimal overhead on performance and decrease the height.

\subsection{Reinforcement Learning-Based Optimization}
\vspace{3pt}
\noindent
\textbf{Background on the Learning Approach.}
Reinforcement Learning (RL) is a machine learning technique that enables an agent to acquire knowledge through trial and error within an interactive environment, driven by feedback from its actions and experiences~\cite{mnih2013playing}. This method has demonstrated significant effectiveness in optimizing system parameters. Q-learning is a popular and efficient RL algorithm that identifies the optimal action based on the current state of the system ($s_t$). The algorithm employs a Q-table to maintain the expected reward values for each possible action in any given state.  At each time step $t$, the agent randomly chooses an action ($a_t$) for which it lacks information (referred to as \textit{exploration}) or picks the action with the highest reward from the Q-table (referred to as \textit{exploitation}).  The balance between exploration and exploitation is managed by the parameter $\epsilon$. Epsilon `decay' can be implemented, progressively decreasing the exploration rate over time to favor exploitation as the agent accumulates more experience. The Bellman equation (Equation~\ref{equation_bellman}) is used to update the Q-table, where $\alpha$ denotes the learning rate and $\gamma$ represents the discount factor that mediates between immediate and future rewards. The reward $r_t$ indicates the result of performing an action in a specific state, essentially showing the effectiveness of the action $a_t$ in state $s_t$. 

\begin{equation}
\label{equation_bellman}
\resizebox{.8\hsize}{!}{$
    Q^{new}(s_t, a_t)=(1-\alpha)Q(s_t, a_t) + \alpha(r_{t+1} + \gamma \underset{a'}max Q(s_{t+1}, a'))
$}
\end{equation}

\noindent Equation \ref{equation_bellman} is utilized to adjust $Q$-Table when we find ourselves in state $s_{t+1}$.

\begin{figure}[t]
  \centering
  \makebox[\columnwidth][c]{\includegraphics[width=\columnwidth]{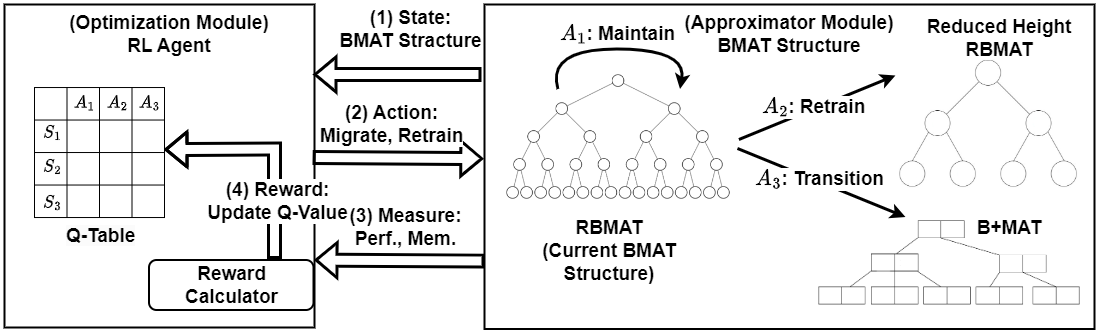}}
  \caption{\small{} RL Agent Overview in four steps. It (1) retrieves the current state (BMAT structure) from the Approximator module, (2) executes the best action which can be (i) keeping the current BMAT structure, (ii) retraining on the subset of data, or (iii) migrating to another BMAT type. (3, 4) The RL agent finally measures the performance and memory consumption of the executed action and updates the Q-Values in the Q-Table based on the calculated reward.}
  \label{fig:rl}
\end{figure}

\vspace{3pt}
\noindent
\textbf{RL Agent.}
We introduce an RL agent to meet the dual objectives of improving system throughput and minimizing memory size. Figure~\ref{fig:rl} shows the overview of the RL agent. This agent facilitates the discovery of optimal operational strategies through continuous interactions within a well-defined environment, described by a state space $\mathcal{S}$, an action space $\mathcal{A}$, and a reward function $R$.

\vspace{3pt}
\noindent
\textbf{States.}
We consider the state space $\mathcal{S}$ to encapsulate critical system metrics and configurations. Each state $s \in \mathcal{S}$ is a vector ($\mathcal{S}_1$, $\mathcal{S}_2$, $\mathcal{S}_3$, $\mathcal{S}_4$, $\mathcal{S}_5$). The four initial state parameters ($\mathcal{S}_1$ to $\mathcal{S}_4$) correspond to the four performance measures outlined in Section~\ref{sec:pm}, while the final parameter specifies the BMAT type.  
\begin{itemize}[label={},leftmargin=*]
    \item $\mathcal{S}_1$: Integer representing the current height of the tree structure.
    \item $\mathcal{S}_2$: Integer representing the minimum granularity of the data coverage by the models at the leaf nodes.
    \item $\mathcal{S}_3$: Real number reflecting the proportional error variation from a baseline, indicative of the stability of the model.
    \item $\mathcal{S}_4$: Integer counts the total number of models active within the system and describes its operational scale.
    \item $\mathcal{S}_5$: Binary variable indicating the type of tree structure, with 0 for RBMAT and 1 for B+MAT.
\end{itemize}

\vspace{3pt}
\noindent
\textbf{Actions.}
The action space $\mathcal{A}$ has discrete actions designed to optimize system configuration based on the actions we defined in Section~\ref{sec:tune}.
\begin{itemize}[label={},leftmargin=*]
    \item $\mathcal{A}_1$: Maintain current BMAT structure.
    \item $\mathcal{A}_2$: Trigger retraining of index models on specific BMAT's branches.
    \item $\mathcal{A}_3$: Transition to another BMAT structure (RBMAT to B+MAT, B+MAT to RBMAT).

\end{itemize}

\vspace{3pt}
\noindent
\textbf{Reward.}
The reward function $R : \mathcal{S} \times \mathcal{A} \rightarrow \mathbb{R}$ is designed to balance throughput and memory usage:
\[
R(s, a) = \eta \cdot \frac{\text{Throughput}(s, a)}{\text{Max System Throughput}}- (1 - \eta) \cdot \frac{\text{Memory Usage}(s, a)}{\text{Total Memory}},
\]
where $\eta$ is the coefficient that weighs the relative importance of throughput and memory efficiency.

\vspace{3pt}
\noindent
\textbf{RL Algorithm.} Algorithm~\ref{alg:tune} shows the steps of the RL agent for learning the best tuning actions at each state using Q-Learning. The agent first observes the current state $s$ and then finds the available actions as the database admin might remove any of the retraining or migrations actions (line~\ref{line:init} to \ref{line:avalid}). Then, it chooses a random action or chooses the best action by comparing the $\epsilon$ value with a random value (line~\ref{alg_eps_begin} to \ref{alg_eps_end}). Then, it tunes the system using the chosen action and runs operations for $N$ times (we choose 1000 in our algorithm) and gets the average time for each operation by calculating the time start from the tuning point to consider the tuning overhead too (line~\ref{line:tune} to \ref{line:measure}). Then, it calculates the reward using the measured throughput and the memory usage of the system and updates the Q-Table using the reward.

\begin{algorithm}[t]
\caption{System Tuning Agent Algorithm}\label{alg:tune}
\begin{algorithmic}[1]
\State $Q \gets \text{initializeQTable()}$ \label{line:init}
\State $\alpha, \gamma, \epsilon \gets \text{initializeVariables()}$ \label{alg:params}
\While {true}
\State \text{Observe state} $s_t$ \label{alg_observe}
\State $A_t \gets \text{getAvailableActions}(s_t)$ \label{line:avalid}
\If {$\text{generateRandNumber()} < \epsilon$} \label{alg_eps_begin}
\State $a_t \gets \text{getRandomAction}(A_t)$
\Else
\State $a_t \gets \arg\max_{a \in A_t} Q(s_t,a)$ \label{alg_eps_end}
\EndIf
\State \text{tuneSystem}($a_t$) \label{line:tune}
\For{$k \gets 1$ to $N$}
\State \text{fetchRunOperation()}
\EndFor
\State \text{Observe state} $s_{t+1}$
\State $P_{t+1}, M_{t+1} \gets \text{measureThroughputMemory()}$ \label{line:measure}
\State $R_{t+1} \gets \text{calculateReward}(P_{t+1},M_{t+1})$
\State $a' \gets \arg\max_{a \in \text{Actions}} Q(s_{t+1},a)$
\State $Q(s_t, a_t) \gets (1-\alpha)Q(s_t, a_t) + \alpha(R_{t+1} + \gamma Q(s_{t+1},a'))$
\State $\epsilon \gets \text{updateEpsilon}(\epsilon)$ \label{alg_update_epsilon}
\State $s_t \gets s_{t+1}$
\EndWhile
\end{algorithmic}
\end{algorithm}

\section{Evaluations}
\label{sec:eval}
In this section, we first outline the experimental setup employed to evaluate UpLIF. Subsequently, we present UpLIF's performance across various metrics and compare it with leading indexing solutions. We demonstrate that UpLIF surpasses both traditional and learned indexes in terms of performance and memory efficiency.

\subsection{Experimental Settings}
\label{sec:exp_setup}
\textbf{Environment.}
We implement UpLIF in C++ and compile it with GCC 9.0.1. We perform our evaluation on an Ubuntu 20.04 Linux machine with AMD Ryzen ThreadRipper Pro 5995WX 64-core 2.7GHz CPU and 256GB DDR4 RAM. All experiments were performed with RadixSpline serving as the underlying learned index model, configured with a spline degree of 128.

\vspace{3pt}
\noindent
\textbf{Datasets.} 
We conduct all experiments using 8-byte keys sourced from a dataset and 8-byte values generated randomly. We evaluated UpLIF on the three following datasets from the SOSD benchmark~\cite{sosd}:
\textit{(1) FB~\cite{FB}:}  200M Facebook user ids, 
\textit{(2) WikiTS~\cite{WikiTS}:}   190M unique request integer timestamps of log entries of the Wikipedia website, and
\textit{(3) Logn:}  200M unique values sampled from a log-normal distribution with heavy tail with $\mu$ = 0 and $\sigma = 1$.

\noindent
For the key-value pairs for index bulk loading, we sort them according to their keys before calling the UpLIF bulk loading.

\vspace{3pt}
\noindent
\textbf{Baselines.} We compare UpLIF's results with the following methods to have comparisons with all the updatable learned index solutions as well as the classical indexing method: \textit{(1) B+Tree:}  This baseline represents a standard B+Tree implementation, implemented in STX B+Tree~\cite{stx}. \textit{(2) Alex \cite{ding2020alex}:} A state-of-the-art learned index method that utilizes an in-place approach for handling updates. \textit{(3) LIPP \cite{10.1145/3589284}:} A state-of-the-art leaned-index using the delta-buffer method for updates. \textit{(4) DILI \cite{10.14778/3598581.3598593}:} A hybrid learned index model that combines different indexing techniques to achieve efficient updates.

\vspace{3pt}
\noindent
\textbf{Workloads.}
The primary measure used to assess UpLIF is the average throughput. Throughput is evaluated across four different workloads: \textit{(1) Read-Only:} a workload focused on read-only operations, \textit{(2) Read-Heavy:} a workload with a high proportion of reads (90\%) and a small percentage of inserts (10\%), \textit{(3) Write-Heavy:} a workload involving 50\% reads and 50\% inserts, and \textit{(4) Write-Only:} a workload dedicated solely to write operations. 

To begin with, an index is initialized with 100 million keys for a given dataset. Each workload is executed for 60 seconds and the remaining keys are inserted during this time. The reported throughput indicates the number of operations (i.e., inserts and reads) completed within that timeframe, averaged over 10 runs to account for variability. 

\vspace{3pt}
\noindent
\textbf{RL Hyperparameters.}
To determine the optimal values for the hyperparameters learning rate ($\alpha$) and discount factor ($\gamma$), we conduct a sensitivity analysis. We consider three classes of values for each parameter: low (0.2), medium (0.5), and high (0.8). RL agent is being trained on Read-Heavy workload on WikiTS using the nine combinations of these values, resulting in a total of nine experiments. Through our analysis, we observe that the agent always performs most effectively when the learning rate is set to a high value and the discount factor to a low value. We also set $\eta=0.7$ in the reward function.

\vspace{3pt}
\noindent
\textbf{RL Training.}
We pre-train the RL agent for each workload and use the trained agent in the Optimization Module (Figure~\ref{frameworkOverview}) to tune the system based on the system condition. In our evaluations, the RL agent only exploits the calculated Q-Table.

\begin{table*}[]
\caption{\small{} Throughput: Comparisons with state-of-the-art methods}
\label{tab:throughput}
\resizebox{\textwidth}{!}{%
\begin{tabular}{c|c|ccccc|cc}
\hline
\multirow{2}{*}{Workload} & \multirow{2}{*}{Dataset} & UpLIF & B+Tree & Alex & LIPP & DILI & \multirow{2}{*}{\begin{tabular}[c]{@{}c@{}}Average\\ Improvement\end{tabular}} & \multirow{2}{*}{\begin{tabular}[c]{@{}c@{}}Max\\ Improvement\end{tabular}} \\
 &  & \multicolumn{5}{c|}{Million Operations Per Second} &  &  \\ \hline
\multirow{3}{*}{\begin{tabular}[c]{@{}c@{}}Read-Only\\ (0\% write rate)\end{tabular}} & WikiTS & \textbf{5.4} & 2.0 & 4.1 & 5.2 & \textbf{5.4} & 50.5\% & 2.68x \\ \cline{2-2}
 & Logn & \textbf{6.3} & 1.9 & \textbf{6.2} & 6.0 & \textbf{6.3} & \textbf{56.1\%} & \textbf{3.18x} \\ \cline{2-2}
 & Facebook & \textbf{4.5} & 2.0 & 2.3 & 4.3 & 4.4 & 55.5\% & 2.23x \\ \hline
\multirow{3}{*}{\begin{tabular}[c]{@{}c@{}}Read-Heavy\\ (10\% write rate)\end{tabular}} & WikiTS & \textbf{4.2} & 1.5 & 2.2 & 3.9 & 4.1 & \textbf{69.3\%} & \textbf{2.78x} \\ \cline{2-2}
 & Logn & \textbf{4.1} & 1.5 & 3.2 & 2.9 & 3.8 & 61.2\% & 2.71x \\ \cline{2-2}
 & Facebook & \textbf{2.6} & 1.3 & 1.1 & 2.3 & 2.5 & 62.8\% & 2.28x \\ \hline
\multirow{3}{*}{\begin{tabular}[c]{@{}c@{}}Write-Heavy\\ (50\% write rate)\end{tabular}} & WikiTS & \textbf{3.4} & 1.3 & 1.9 & 2.8 & 2.1 & 73.8\% & 2.61x \\ \cline{2-2}
 & Logn & \textbf{3.9} & 1.3 & 3.2 & 2.9 & 2.9 & 76.2\% & \textbf{3.00x} \\ \cline{2-2}
 & Facebook & \textbf{3.9} & 1.3 & 1.1 & 2.1 & 2.4 & \textbf{82.2\%} & 2.48x \\ \hline
\multirow{3}{*}{\begin{tabular}[c]{@{}c@{}}Write-Only\\ (100\% write rate)\end{tabular}} & WikiTS & \textbf{2.8} & 1.3 & 1.4 & 2.2 & 2.4 & 66.2\% & 2.14x \\ \cline{2-2}
 & Logn & \textbf{3.1} & 1.2 & 2.8 & 2.2 & 2.7 & 57.6\% & \textbf{2.36x} \\ \cline{2-2}
 & Facebook & \textbf{2.5} & 1.2 & 1.2 & 1.7 & 1.9 & \textbf{68.1\%} & 2.04x \\ \hline
\multirow{3}{*}{\begin{tabular}[c]{@{}c@{}}Distribution Shift\\ (50\% write rate)\end{tabular}} & WikiTS & \textbf{2.7} & 1.3 & 1.7 & 2.2 & 1.8 & 54.8\% & 2.08x \\ \cline{2-2}
 & Logn & \textbf{3.1} & 1.3 & 2.8 & 2.3 & 2.2 & \textbf{59.4\%} & \textbf{2.49x} \\ \cline{2-2}
 & Facebook & \textbf{2.1} & 1.3 & 1.0 & 1.6 & 1.4 & 59.0\% & 1.93x \\ \hline
\end{tabular}
}

\end{table*}
\vspace{-3pt}
\subsection{Performance on the Workloads}
Table~\ref{tab:throughput} shows the performance of UpLIF in terms of throughput compared to the other baselines in different workloads and datasets. Our results indicate that as the update rate increases in workload, the gap in throughput improvement between UpLIF and other learned index models widens. This can be attributed to UpLIF's optimization for updates through model adjustment with minimal retraining (in the Approximator module) as well as the index structure tuning (in the Optimization Agent module). As the number of updates increases, other methods such as DILI require more frequent retraining and node conflict handling, resulting in lower throughput. In contrast, UpLIF's efficient update handling allows it to maintain higher throughput even with a high update rate, further highlighting its superiority in handling dynamic workloads.

\vspace{3pt}
\noindent
\textbf{Read-Only Workloads.}
In the Read-Only workload scenario, we train UpLIF on the entire dataset without any updates. In this case, UpLIF's performance is identical to the underlying model it utilizes (i.e., RadixSpline).  UpLIF consistently achieves more than $2\times$ throughput compared to B+Tree in all datasets. However, compared to the other learned index frameworks, UpLIF's throughput is comparable to DILI since no updates have been applied to UpLIF in this particular workload. In subsequent workloads where the number of updates increases, we observe that UpLIF consistently outperforms all other learned index models, resulting in higher throughput.

\vspace{3pt}
\noindent
\textbf{Read-Heavy Workloads.}
For Read-Heavy workloads, UpLIF begins to lead all baselines in performance as there are a small number of updates in this workload type. UpLIF achieves up to $2.78\times$ throughput compared to B+Tree on the Logn Dataset. UpLIF also outperforms all other learned index models and achieves more than $60\%$ throughput on average. Although DILI has less performance than UpLIF, it performs relatively well as the number of updates is low, and its structure is not influenced by the updates.

\vspace{3pt}
\noindent
\textbf{Write-Heavy Workloads.}
In the Write-Heavy workload, UpLIF exhibits its best performance due to its model adjustment and structure tuning by retraining and switching between BMAT trees. Compared to B+Tree, UpLIF shows up to $3\times$ higher throughput, and compared to the state-of-the-art updatable learned indexes, it achieves up to 82\% higher performance. Specifically, considering the Facebook dataset, UpLIF achieves 3.9 million operations per second, while B+Tree, Alex, LIPP, and DILI are far behind with 1.3, 1.1, 2.1, and 2.4 million ops/sec.

\vspace{3pt}
\noindent
\textbf{Write-Only Workloads.}
The Write-Only workload is the most challenging since the index structure is being influenced the most. However, the tuning and approximation techniques of the UpLIF structure mitigate this impact and outperform all other benchmarks. UpLIF achieves performance levels up to $2.49\times$ greater than B+Tree. Furthermore, it delivers an average throughput increase of up to 59.4\% in all three datasets compared to the learned indexes.

\vspace{-3pt}
\subsection{Distribution Shift}
To evaluate UpLIF and the distribution shift baselines, we initialize the keys with the first 100 million smallest keys for each dataset and then use the write-heavy workload to read and write the remaining data. The last row of Table~\ref{tab:throughput} shows the robustness of UpLIF to changes in the data distribution. UpLIF maintains up to $2.49\times$ and $1.76\times$ higher throughput compared to B+Tree and DILI respectively. DILI cannot work well on distribution shift, as its structure is built based on the dataset characteristics, and inserting keys from different distributions causes more conflicts in its leaf structures which need more node adjustment.

\vspace{-3pt}
\subsection{Range Query Performance}
\begin{figure}[t]
  \centering
  \makebox[\columnwidth][c]{\includegraphics[width=\columnwidth]{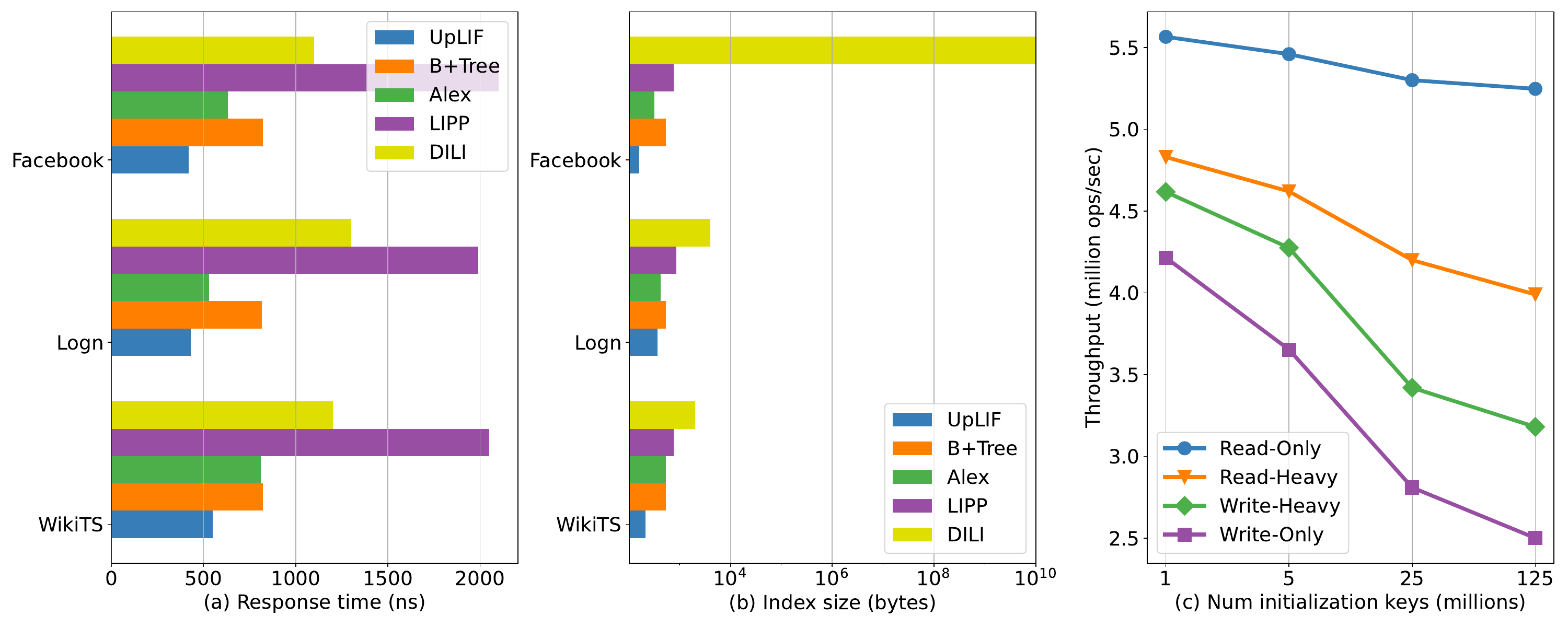}}
  \caption{\small{} (a) Range query performance (b) Index memory sizes (c) Throughput with different numbers of initialization data}
  \label{fig:combie}
\end{figure}

On the range query workload, UpLIF shows significantly lower response times compared to other methods in all data sets (see Figure \ref{fig:combie}a).  This is attributed to UpLIF's optimization for range queries by maintaining sorted data at the leaf nodes. UpLIF outperforms DILI by a considerable margin ranging from 55\% to 70\%. The performance gap between UpLIF and DILI becomes more pronounced due to DILI's less efficient handling of range queries, as its model tree necessitates frequent retraversals, resulting in slower processing and longer response times. Similarly, other methods such as B+Tree, Alex, and LIPP also exhibit higher response times compared to UpLIF. For instance, B+Tree and Alex consistently demonstrate slower performance by notable margins, while LIPP exhibits the highest response times among all the tested methods. These comparisons underscore UpLIF's superior efficiency and reliability in efficiently processing range queries, establishing it as the most effective solution across all the evaluated datasets.

\vspace{-3pt}

\subsection{Index Memory Consumption}
We compare the memory usage of the index structure in UpLIF under a write-heavy workload across different datasets and compare it with other baselines. Figure~\ref{fig:combie}b shows that UpLIF always achieves significantly lower memory consumption compared to the other methods, with a notable difference of up to 1000 times lower memory usage compared to DILI. This advantage is primarily due to UpLIF's ability to determine the update distribution and strategically place placeholders to efficiently collect updates, thereby minimizing the need to add nodes to BMAT. In contrast, DILI and LIPP exhibit the highest memory consumption, since these models create new leaf nodes and empty slots in the entry array whenever conflicts occur. Although Alex utilizes a similar in-place placeholder approach as UpLIF, it still results in slightly higher memory consumption due to its inefficiency in accurately placing the placeholders based on the update distribution.

\subsection{Scalability}
Figure \ref{fig:combie}c shows the UpLIF's throughput scalability on different scales of initialization keys, ranging from 1 million to 125 million using the WikiTS dataset. The diagram organizes the data according to the four workloads described earlier. As observed, in the Read-Only and Read-Heavy scenarios, the system maintains a high throughput, demonstrating its efficiency in handling large volumes of read operations. The throughput decreases slightly but remains robust even as the initialization keys increase. In Write-Heavy and Write-Only workloads, initially, there is a noticeable decline in throughput as the volume of initial data expands. However, once a specific threshold of initial keys is reached, the BMAT begins to retrain and prune the tree, which stabilizes the system's throughput.

\section{Conclusion}

UpLIF introduces an efficient and robust updatable learned index framework that overcomes the limitations of traditional indexes. It incorporates the Balanced Model Adjustment Tree and uses a hybrid approach with placeholders and delta buffers to manage updates without frequent retraining. It also incorporates an RL-based system tuning to optimize its index structure. Our extensive testing shows that UpLIF exceeds other advanced indexing methods in various workloads and maintains high performance and low memory usage even with distribution changes and increases in the data scale.

\bibliographystyle{splncs04}
\bibliography{uplif}

\end{document}